\documentstyle[12pt]{article}
\def\bea{\begin{eqnarray}}
\def\eea{\end{eqnarray}}
\def\nn{\nonumber}

\def\beq{\begin{equation}}
\def\eeq{\end{equation}}
\def\ba{\beq\new\begin{array}{c}}
\def\ea{\end{array}\eeq}
\def\be{\ba}
\def\ee{\ea}
\def\stackreb#1#2{\mathrel{\mathop{#2}\limits_{#1}}}
\def\Tr{{\rm Tr}}
\def\res{{\rm res}}

\makeatletter
\newdimen\normalarrayskip              % skip between lines
\newdimen\minarrayskip                 % minimal skip between lines
\normalarrayskip\baselineskip
\minarrayskip\jot
\newif\ifold             \oldtrue            \def\new{\oldfalse}
\def\arraymode{\ifold\relax\else\displaystyle\fi} % mode of array entries
\def\eqnumphantom{\phantom{(\theequation)}}     % right phantom in eqnarray
\def\@arrayskip{\ifold\baselineskip\z@\lineskip\z@
     \else
     \baselineskip\minarrayskip\lineskip2\minarrayskip\fi}
\def\@arrayclassz{\ifcase \@lastchclass \@acolampacol \or
\@ampacol \or \or \or \@addamp \or
   \@acolampacol \or \@firstampfalse \@acol \fi
\edef\@preamble{\@preamble
  \ifcase \@chnum
     \hfil$\relax\arraymode\@sharp$\hfil
     \or $\relax\arraymode\@sharp$\hfil
     \or \hfil$\relax\arraymode\@sharp$\fi}}
\def\@array[#1]#2{\setbox\@arstrutbox=\hbox{\vrule
     height\arraystretch \ht\strutbox
     depth\arraystretch \dp\strutbox
     width\z@}\@mkpream{#2}\edef\@preamble{\halign
\noexpand\@halignto
\bgroup \tabskip\z@ \@arstrut \@preamble \tabskip\z@ \cr}%
\let\@startpbox\@@startpbox \let\@endpbox\@@endpbox
  \if #1t\vtop \else \if#1b\vbox \else \vcenter \fi\fi
  \bgroup \let\par\relax
  \let\@sharp##\let\protect\relax
  \@arrayskip\@preamble}
\def\eqnarray{\stepcounter{equation}%
              \let\@currentlabel=\theequation
              \global\@eqnswtrue
              \global\@eqcnt\z@
              \tabskip\@centering
              \let\\=\@eqncr
              $$%
 \halign to \displaywidth\bgroup
    \eqnumphantom\@eqnsel\hskip\@centering
    $\displaystyle \tabskip\z@ {##}$%
    \global\@eqcnt\@ne \hskip 2\arraycolsep
         $\displaystyle\arraymode{##}$\hfil
    \global\@eqcnt\tw@ \hskip 2\arraycolsep
         $\displaystyle\tabskip\z@{##}$\hfil
         \tabskip\@centering
    &{##}\tabskip\z@\cr}

\begin{document}

\setcounter{footnote}0
\begin{center}
\hfill FIAN/TD-11/96,\ ITEP/TH-23/96\\
%\hfill \today\\
\vspace{0.3in}
{\LARGE\bf From Nonperturbative SUSY Gauge Theories to
Integrable Systems
\footnote{based on a talk given at 10th International Conference
{\sl Problems of Quantum Field Theory}, 13-18 May 1996, Alushta, Ukraine}}\\
\bigskip\bigskip
{\Large A.Marshakov}
\footnote{E-mail address:
mars@lpi.ac.ru}\\
{\it Theory Department,  P. N. Lebedev Physics
Institute , Moscow,~117924, Russia}\\ and {\it ITEP, Moscow 117259, Russia}\\
\end{center}
\bigskip\bigskip

\begin{abstract}
I review the appearence of integrable structures
in the formulation of exact nonperturbative solutions to $4d$
supersymmetric quantum
gauge theories. Various examples of ${\cal N}\geq 2$ SUSY Yang-Mills
nonperturbative solutions are adequately described in terms of
the (deformations of the) finite-gap solutions to integrable models:
through the generating differential and the $\tau$-function.
One of the basic definitions of generating differential is discussed and
its role in the theory of integrable systems is demonstrated.
\end{abstract}

{\bf 1. INTRODUCTION}.
Recent investigations showed that not only the simplest (topological)
string model's generating functions but also the
low-energy effective actions and the BPS massive
spectra for $4d$ (at least) ${\cal N}\geq 2$ SUSY Yang-Mills  theories
\cite{SW} possess a nice
description in terms of the integrable systems \cite{GKMMM}.
The relation between the Seiberg-Witten solutions (SW) and
integrable theories
is already established in detail for two families of
models: the ${\cal N}=2$ SYM theory with one ($N_a=1$)
``matter" ${\cal N}=2$ hypermultiplet in the adjoint representation
-- related to the
Calogero-Moser integrable models \cite{Cal}
and to the ${\cal N}=2$ SYM QCD -- theory with ($N_f$) fundamental
matter hypermultiplets -- connected with the family of integrable spin chains
\cite{Spin}
\footnote{It is known also \cite{Mar},\cite{AN} that the
$N_c = 3$, $N_f = 2$ curve can be associated with the
Goryachev-Chaplygin top.}.

The SW solution can be formally defined as a map
\be
G,\tau,h_i \rightarrow T_{ij}, \ a_i,\ a_i^D
\ee
and the solution to this problem has an elegant description in the following
terms \cite{SW}:
one associates with the data $G$ (gauge group), $\tau$ (the UV coupling
constant), $h_k$ ($={1\over k}\langle\Tr\Phi ^k\rangle$ -- the v.e.v.'s
of the Higgs fields) a family of complex curves $\Sigma $ with $h_i$
parameterizing (some) moduli
of their complex structures, and a meromorphic 1-form
$dS$ on every $\Sigma $. The periods
$a_i = \oint_{A_i} dS$, $a_i^D = \oint_{B^i} dS$ determine the BPS massive
spectrum, $\frac{\partial{\cal F}}{\partial a_i}= a_i^D$ the low-energy
effective action ${\cal F}$ and the set of low-energy coupling constants
$T_{ij} = \frac{\partial^2{\cal F}}{\partial a_i\partial a_j} =
\frac{\partial a^D_i}{\partial a_j}$.

In terms of integrable systems the curves $\Sigma $ are
interpreted \cite{GKMMM} as spectral curves of certain
integrable systems, and $a_i$, $a_i^D$ are related to the action integrals
($\oint pdq$) of the system. To describe the solution one should present the
explicit construction
\be
G, \tau, h_k \rightarrow \left( \Sigma ,\ dS
\right)\{h_i\},
\ee
and this turns to be equivalent to a selection of particular integrable
model.

{\bf 2. CURVES AND INTEGRABLE SYSTEMS}. First let us discuss the {\it fact} of
appearence of (finite-dimensional)
integrable systems in the framework of the quantum $4d$ field theories.
Indeed, since the solutions are formulated in terms of periods of some
differential on a complex curve -- it means that an integrable system
(moreover an integrable system of KP/Toda type)
arises more or less by definition. The arrow from curves to the finite-gap
solutions is provided by the Krichever construction, while the arrow in
the opposite direction by the Novikov hypothesis proven by Shiota. Since
till the moment no real progress has been achieved in deducing the integrable
equations directly from the basic definitions of the quantum field (string)
theory -- this relation can be considered just as an observation. Below in
this section, reviewing the main statements of refs. \cite{Cal,Spin}, I
stress that in fact this is a {\it useful} observation leading to the
possibility of applying rather simple technique of $\leq 2$-dimensional
integrable
systems (the Lax pairs, spectral curves, symplectic forms and Hamiltonian flows,
$\tau $-functions, etc) to $4d$ quantum field theories.

To be more concrete, let us start with the observation \cite{GKMMM} that
the nonperturbative solution to the pure ${\cal N}=2$ SYM theory with $SU(N_c)$
gauge groups is described in terms of the periodic Toda-chain spectral
curves.
The periodic problem in this model can be formulated in two different elegant
ways, which can be {\it naturally deformed} in two different directions.
The useful observation is that exactly these two deformations
are related to the two {\it physically interesting} deformations
of the $4d$ theory by coupling it to the adjoint and fundamental matter
hypermultiplets.

The Toda-chain system is defined by the equations of motion with the pairwise
exponential interaction
\be
\frac{\partial q_i}{\partial t} = p_i \ \ \ \ \
\frac{\partial p_i}{\partial t} = e^{q_{i+1} -q_i}- e^{q_i-q_{i-1}}
\ee
and the $N_c$-periodic
problem, corresponding to the situation when one has exactly $N_c$ particles
living on a circle
(that just means that $N_c$-th particle interacts as well with the first one)
can be equivalently formulated in terms of two different Lax representations.
The spectral curve itself arises as a condition of the common spectrum of the
Lax operator ${\cal L}$ and the shift operator $T_{N_c}$ responsible for the
periodic boundary conditions.
In the first version the Lax operator of the auxiliary linear problem
\be\label{laxtoda}
\lambda\psi ^{\pm}_n =
\sum _k {\cal L}_{nk}\psi ^{\pm}_k =
e^{{1\over 2}(q_{n+1}-q_n)}\psi ^{\pm}_{n+1} + p_n\psi ^{\pm}_n +
e^{{1\over 2}(q_n-q_{n-1})}
\psi ^{\pm}_{n-1}
\ (=  \pm {\partial\over\partial t}\psi ^{\pm}_n)
\ee
becomes  the
$N_c\times N_c$ matrix-valued 1-form \cite{KriDu} -- that corresponds to the
fact that it
is defined on the {\it eigenvectors} of the shift operator $T_{N_c}$,
and the eigenvalues of the Lax operator are defined from the spectral
equation
\be\label{fsc-Toda}
\det_{N_c\times N_c}\left({\cal L}^{TC}(w) -
\lambda\right) = 0 \ \
\Rightarrow
w + \frac{1}{w} = 2P_{N_c}(\lambda )
\ee
where $w$ is the eigenvalue of the $T_{N_c}$-operator and $P_{N_c}(\lambda )$
-- a polynomial of degree $N_c$ with the coefficients being the symmetric
functions of the Toda chain Hamiltonians -- identified with the v.e.v.'s
of the Higgs field $h_k$.

An alternative description of the same system arises when one {\it solves}
explicitely the auxiliary linear problem (\ref{laxtoda}) which is just a
{\it second}-order
difference equation and rewrites the solution ${\tilde\psi}_{i+1}
= L^{TC}_i(\lambda ){\tilde\psi}_i$ (after a simple "gauge"
transformation) with the help of a chain of $2\times 2$ Lax matrices \cite{FT}
\be\label{LTC}
L^{TC}_i(\lambda) =
\left(\begin{array}{cc} p_i + \lambda & e^{q_i} \\ e^{-q_i} & 0
\end{array}\right), \ \ \ \ \ i = 1,\dots ,N_c
\ee
manifestly depending on the Lax eigenvalue -- in contrast to the previous
picture now we work with the Lax eigenfunctions. The shift operator becomes
\be\label{monomat}
T_{N_c}(\lambda) =
\prod_{N_c\leq i\leq 1} L_i(\lambda )
\ee
and the (same!) curve and integrals of motion of the Toda chain are generated
by {\it another} form of spectral equation (with $2\times 2$ instead of
$N_c \times N_c$ matrices)
\be\label{specTC}
\det_{2\times 2}\left( T^{TC}_{N_c}(\lambda )
- w\right) = w^2 - w\Tr T^{TC}_{N_c}(\lambda ) + 1 = 0\ \
\Rightarrow
w + \frac{1}{w} = \Tr T^{TC}_{N_c}(\lambda) \equiv 2P_{N_c}(\lambda )
\ee
The generating 1-form (to be discussed in detail in the next section)
\be\label{dS}
dS \cong \lambda d\log w = \Tr{\cal L}d\log T_{N_c}
\ee
is defined by the eigenvalues of two (commuting on a spectral curve)
operators.

The $N_c\times N_c$ matrix-valued
Lax 1-form comes naturally from the $GL(N_c)$ Calogero system \cite{KriCal}
${\cal L}(\xi)d\xi$ which is
defined on elliptic curve $E(\tau)$.
The Calogero coupling constant in $4d$
interpretation plays the role of the mass of the adjoint matter
${\cal N}=2$ hypermultiplet breaking ${\cal N}=4$ SUSY down to ${\cal N}=2$
\cite{Cal}.
The spectral curve $\Sigma ^{Cal}$ for the Calogero system:
\be\label{fscCal}
\det_{N_c\times N_c} \left({\cal L}^{Cal}(\xi) - \lambda\right) = 0
\ee
and the periods $a_i$ and $a_i^D$ of the
generating 1-differential
\be\label{dSCal}
dS^{Cal} \cong \lambda d\xi
\ee
in the double-scaling limit
($x_i^{Cal}-x_j^{Cal} \sim \left[(i-j)\log g^{Cal} +(q_i-q_j)\right]
\rightarrow\infty$) \cite{Ino}, recover the Toda-chain data (\ref{fsc-Toda})
and (\ref{dS}).
In this limit,
the elliptic curve $E(\tau)$ degenerates into the (two-punctured) Riemann
sphere with coordinate $w = e^{\xi }e^{i\pi\tau}$ so that
\be
dS^{Cal} \rightarrow dS^{TC} \cong \lambda\frac{dw}{w}
\ee
In contrast to the Toda case, $\Sigma ^{Cal}$ (\ref{fscCal}) can {\it not} be
rewritten in
the form (\ref{fsc-Toda}) and the specific $w$-dependence of the spectral
equation (\ref{fsc-Toda}) is not preserved by embedding of Toda into
Calogero-Moser
system. However, the form (\ref{fsc-Toda}) {\it is preserved} by
the alternative deformation of the Toda-chain system when considering
it as (a particular case of) a spin-chain model.

The full spectral curve for the periodic spin chain is
given by:
\be\label{fsc-SCh}
\det_{2\times 2}\left(T_{N_c}(\lambda) -  w\right) = 0,
\ee
with the inhomogeneous $T$-matrix
\be\label{T-matrix}
T_{N_c}(\lambda) = \prod_{i=N_c}^1 L_i(\lambda-\lambda_i)
\ee
(in general $\det T_{N_c}(\lambda ) \equiv Q(\lambda ) \neq 1$!)
and generating differential is now
\be\label{1f}
dS = \lambda\frac{d\tilde w}{\tilde w} \ \ \ \ \
{\tilde w} = w\cdot (\det T_{N_c})^{-1/2}
\ee
or
\be\label{fsc-sc1}
w + \frac{Q(\lambda)}{w} =
 2P_{N_c}(\lambda ) \ \ \ \ \
{\tilde w} + \frac{1}{\tilde w} = \frac{2P_{N_c}(\lambda)}
{\sqrt{Q(\lambda)}}
\ee
which is a proposed form of a curve for ${\cal N}=2$ SUSY QCD.

{\bf 3. SYMPLECTIC FORM}.
Now let us turn to the discussion of a more subtle point -- why the
generating 1-form (\ref{dS}) indeed describes an integrable system.
To do this I will discuss the symplectic structure on the space of the
finite-gap solutions.
This symplectic structure was introduced in
\cite{DKN} and recently proposed in \cite{KriPho} as coming directly
from the symplectic form on the space of all the solutions to the hierarchy.
Below, a very simple and straghtforward proof
of this result is presented and the relation with the analogous object in
low-dimensional non-perturbative string theory \cite{KM} is discussed.

To prove that (\ref{dS}) is a generating one-form of the whole hierarchy
one starts with the variation of the generating function
\be\label{S}
S(\Sigma ,{\bf \gamma}) = \sum _i \int ^{\gamma _i}Edp
\ee
(where $dE$ and $dp$ ($= d\lambda $ and $={dw\over w}$ in the particular case
above) are two meromorphic
differentials on a spectral curve $\Sigma $ and $\bf \gamma $ is the divisor
of the solution (poles of the BA function))
\be\label{var}
\delta S = \sum _i (Edp)(\gamma _i) + \sum _i \int ^{\gamma _i}\delta E dp
\ \ \ \ \ \
\delta ^2 S = \delta\left( \sum _i (Edp)(\gamma _i)\right)
+ \sum _i (\delta E dp)(\gamma _i)
\ee
From $\delta ^2 S = 0$ it follows that
\be\label{wsform}
\varpi  = \delta E\wedge\delta p =
\delta\left( \sum _i (Edp)(\gamma _i)\right) =
- \sum _i (\delta E dp)(\gamma _i)
\ee
Now, the variation $\delta E$ (for constant $p$) follows from the Lax
equation (auxiliary linear problem)
\be\label{lax}
{\partial\over\partial t}\psi = {\cal L}\psi \ (= E\psi )
\ee
so that
\be\label{varE}
\delta E = {\langle \psi ^{\dagger}\delta {\cal L}\psi\rangle\over
\langle \psi ^{\dagger}\psi\rangle}
\ee
and one concludes that
\be\label{general}
\varpi =  - \langle \delta{\cal L}\sum _i \left( dp{\psi ^{\dagger}\psi\over
\langle \psi ^{\dagger}\psi\rangle}\right) (\gamma _i) \rangle
\ee
Let us turn to several important examples.\\
{\bf KP/KdV}. In the KP-case the equation (\ref{lax}) looks as
\be\label{laxkp}
{\partial\over\partial t}\psi = \left({\partial ^2\over\partial x^2} +
u\right) \psi \ (= E\psi )
\ee
therefore the equation (\ref{general}) implied by
$\langle \psi ^{\dagger}\psi\rangle = \int _{dx}\psi ^{\dagger}(x,P)\psi (x,P)$
and $\delta{\cal L} = \delta u(x)$ gives
\be\label{kpcase}
\varpi  =  - \int _{dx} \delta u(x) \sum _i
\left( {dp\over\langle \psi ^{\dagger}\psi\rangle}\psi ^{\dagger}(x)\psi (x)
\right) (\gamma _i)
\ee
The differential
$d\Omega = {dp\over\langle \psi ^{\dagger}\psi\rangle}
\psi ^{\dagger}(x)\psi (x)$ is
holomorphic on $\Sigma $ except for the ''infinity" point $P_0$ where it has
zero residue \cite{KriUMN}. Its variation
\footnote{It should be pointed out that the variation $\tilde\delta $
corresponds to a rather specific situation when one shifts only $\psi $
keeping $\psi ^{\dagger}$ fixed.}
\be\label{reskp}
{\tilde\delta}\left(\res _{P_0}d\Omega + \sum _i\res _{\gamma _i}d\Omega \right) = 0
\ee
can be rewritten as
\be\label{reskp2}
\delta v(x) + \sum _i d\Omega (\gamma _i) = 0
\ee
where $v(x)$ is a ''residue" of the BA function at the point $P_0$ obeying
$v'(x) = u(x)$. Substituting (\ref{reskp2}) into (\ref{general}) one gets
\be\label{kdv1}
\varpi = \int _{dx} \delta u(x) \int ^x _{dx'}\delta u(x')
\ee
or the {\it first} symplectic structure of the KdV equation.\\
{\bf Toda chain/lattice}. (The case directly related to the {\it
pure} SYM theory). One has
$\langle \psi ^{\dagger}\psi\rangle = \sum _n\psi ^+_n(P)\psi ^-_n(P)$,
and the Lax equation acquires the form (\ref{laxtoda})
where $t = t_+ + t_-$ and $t_1 = t_+ - t_-$ is the first time of the Toda
chain, so that
\be\label{varEtoda}
\delta\lambda = {\sum _k \psi ^+_k \delta
p_k\psi ^-_k\over \langle \psi ^+\psi ^-\rangle}
\ee
and (\ref{general})
becomes
\be\label{toda}
\varpi  =  - \sum _k \delta p_k \sum _i \left(
{dp\over\langle \psi ^+\psi ^-\rangle}\psi ^+_k \psi ^-_k
\right) (\gamma _i)
\ee
and to get
\be\label{ochain}
\varpi = \sum _k \delta p_k\wedge\delta q_k
\ee
one has to prove
\be\label{varx}
\sum _i \left(
{dp\over\langle \psi ^+\psi ^-\rangle}\psi ^+_k \psi ^-_k
\right) (\gamma _i)  = \delta q_k
\ee
To do this one considers again
\be\label{restoda}
{\tilde\delta} \left(\res _{P_+} + \res _{P_-} +
\sum _i\res _{\gamma _i}\right)d\Omega _n = 0
\ \ \ \ \ \
d\Omega _n =
{dp\over\langle \psi ^+\psi ^-\rangle}\psi ^+_n \psi ^-_n
\ee
where the first two terms for $\psi ^{\pm}_n \stackreb{\lambda
\rightarrow\lambda (P_{\pm})}{\sim} e^{\pm q_n}\lambda ^{\pm n}(1 +
{\cal O}(\lambda ^{-1}))$ satisfying two ''shifted" equations (\ref{laxtoda})
(with
${\tilde q}_n$ and $q_n $ correspondingly) give $\delta q_n = {\tilde q}_n -
q_n$ while the rest -- the l.h.s. of (\ref{varx}).

{\bf Calogero-Moser system}.
Introducing the ''standard" $dE$ and $dp$ one the curve $\Sigma $
(\ref{fscCal})
with the 1-form (\ref{dSCal})
where $dp=d\xi$ is holomorphic on torus $\oint _A dp = \omega$,
$\oint _B dp = \omega '$ and $E=\lambda $
has $n-1$ poles with $residue = 1$ and $1$ pole with $residue = -(n-1)$,
the BA function is defined by \cite{KriCal}
\be\label{bacal}
{\cal L}^{Cal}(\xi ){\bf a} =  \lambda {\bf a}
\ee
with the essential singularities
\be\label{baprop}
a_i \stackreb{E = E_+}{\sim} e^{x_i\zeta (\xi )}\left( 1 + {\cal O}(\xi)\right)
\ \ \ \ \
a_i \stackreb{E \neq E_+}{\sim} e^{x_i\zeta (\xi)}\left( -{1\over n-1} +
{\cal O}(\xi)\right)
\ee
and (independent of dynamical variables) poles ${\bf \gamma}$.  Hence,
similiarly to the above case for the eq. (\ref{general}) one has
$\langle \psi ^{\dagger}\psi\rangle = \sum _ia_i^{\dagger}(P) a_i(P)$,
$\delta {\cal L}^{Cal} = {\sum _ia_i^{\dagger}(P) \delta p_i a_i(P)\over
\sum _ia_i^{\dagger}(P) a_i(P)}$ so that
\be\label{calo}
\varpi =  - \sum _k \delta p_k \sum _i \left(
{d\xi\over\langle a^{\dagger}a\rangle}a^{\dagger}_k a_k
\right) (\gamma _i)
\ee
and the residue formula
\be\label{rescal}
{\tilde\delta} \left(\sum _{P_j:p = 0}\res _{P_j} +
\sum _i\res _{\gamma _i}\right)d\Omega _k = 0
\ \ \ \ \ \ \
d\Omega _k =
{d\xi\over\langle a^{\dagger}a\rangle}a^{\dagger}_k a_k
\ee
where the first sum is over all ''infinities" $p = 0$ at each sheet of the
cover (\ref{fscCal}). After variation and using (\ref{baprop}) it gives
again
\be\label{calogero}
\varpi = \sum _k \delta p_k\wedge \delta x_k
\ee
The general proof of the more cumbersome analog of the above derivation
can be found in \cite{KriPho}. To show how the above
formulas
work explicitely, let us, finally, demonstrate the existence of (\ref{reskp}),
(\ref{restoda}) and (\ref{rescal}) for the 1-gap solution. Let
\be\label{psya}
\psi = e^{x\zeta (z)}
{\sigma (x-z+\kappa)\over\sigma (x+\kappa)\sigma (z-\kappa)}
\ \ \ \  \ \ \ \
\psi ^{\dagger}= e^{-x\zeta (z)}
{\sigma (x+z+\kappa)\over\sigma (x+\kappa)\sigma (z+\kappa)}
\ee
be solutions to
\be\label{1gap}
(\partial ^2 + u)\psi = (\partial ^2 - 2\wp (x+\kappa))\psi = \wp (z)\psi
\ee
Then
\be
\psi ^{\dagger}\psi = {\sigma (x-z+\kappa )\sigma (x+z+\kappa )\over\sigma ^2(x+\kappa )
\sigma (z-\kappa )\sigma (z+\kappa )}
= {\sigma ^2(z)\over\sigma (z +\kappa )\sigma (z - \kappa )}
\left( \wp (z) - \wp (x+\kappa )\right)
\nn \\
\langle \psi ^{\dagger}\psi\rangle =
= {\sigma ^2(z)\over\sigma (z +\kappa )\sigma (z - \kappa )}
\left( \wp (z) - \langle\wp (x+\kappa )\rangle \right)
\ee
and let us take the average over a period $2{\tilde\omega}$ to be
$\langle\wp (x+\kappa )\rangle = 2{\tilde\eta}$. Also
\be\label{qm}
dp = d\left(\zeta (z) + \log\sigma (2{\tilde\omega}-z+\kappa ) - \log\sigma
(\kappa - z)\right) = -dz\left(\wp (z) + \zeta (2{\tilde\omega}-z+\kappa ) -
\zeta (\kappa -z)\right) =
\nn \\
= -dz\left( \wp (z) - 2{\tilde\eta}\right)
\ee
and
\be\label{ppsi}
{dp\over\langle\psi ^{\dagger}\psi\rangle} =
dz{\sigma (z+\kappa )\sigma (z-\kappa)\over\sigma ^2(z)}
\nn \\
d\Omega = {dp\over\langle\psi ^{\dagger}\psi\rangle}\psi ^{\dagger}\psi =
dz{\sigma (x+z+\kappa )\sigma (x+z-\kappa)\over\sigma ^2(z)\sigma ^2(x+\kappa )} =
dz\left(\wp (z) - \wp (x+\kappa )\right)
\ee
Now, the variation $\tilde\delta$ explicitely looks as
\be\label{var1gap}
\tilde\delta d\Omega \equiv {dp\over\langle\psi ^{\dagger}\psi\rangle}
\psi ^{\dagger}_{\kappa}\psi _{\kappa + \delta\kappa} =
dz
{\sigma (z+\kappa )\sigma (z-\kappa)\sigma (x+z+\kappa)
\sigma (x-z+\kappa +\delta\kappa )
\over
\sigma ^2(z)\sigma (x+\kappa)\sigma (z+\kappa)
\sigma (x+\kappa +\delta\kappa )\sigma (z-\kappa -\delta\kappa )} =
\nn \\
= dz{\sigma (x+\kappa +z)\sigma (x+\kappa -z)\over\sigma ^2(z)\sigma ^2(x +\kappa )}
\left[ 1 + \delta\kappa\left(\zeta (x-z+\kappa ) + \zeta (z-\kappa ) -
\zeta (x +\kappa )\right) + {\cal O}\left( (\delta\kappa)^2\right)\right]
\nn \\
= dz\left(\wp (z) - \wp (x +\kappa )\right)
\left[ 1 + \delta\kappa\left(\zeta (x-z+\kappa ) + \zeta (z-\kappa ) -
\zeta (x +\kappa )\right) + {\cal O}\left( (\delta\kappa)^2\right)\right]
\ee
It is easy to see that (\ref{var1gap}) has non-zero residues at $z=0$ and
$z = \kappa $ (the residue at $z = x+\kappa $ is suppressed by
$\wp (z) - \wp (x +\kappa )$. They give
\be\label{res0}
\res _{z=0} \delta d\Omega \sim \delta\kappa \oint _{z\hookrightarrow 0}
dz\wp (z)
\left(\zeta (x-z+\kappa )+\zeta (z-\kappa )\right) \sim
\sim\delta\kappa \oint _{z\hookrightarrow 0}\zeta (z) d\left(\zeta (x-z+\kappa )
+ \right.
\nn \\
\left. +\zeta (z-\kappa )\right)\sim
\delta\kappa\left(\wp (x +\kappa ) +\wp (\kappa )\right)\sim
\delta\left(\zeta (x +\kappa ) + \zeta (\kappa)\right)\equiv \delta v(x)
\ee
and
\be
\res _{z=\kappa }\delta d\Omega = \delta\kappa \left(\wp (\kappa )-\wp (x+\kappa )\right)
= d\Omega (\kappa )
\ee
which follows from the comparison to (\ref{ppsi}).

{\bf 4. CONCLUSION}.
The quantization of the symplectic form (\ref{wsform}) is known
to correspond to the complete description of the effective theory
(not only its low-energy part) at least in the simplest case when
$E = W(\mu )$ and $p=Q(\mu )$ were two functions (polynomials) on a
complex sphere. The corresponding generating function (\ref{dS})
was essential in the definition of the duality transformation between two
dual points with completely different behaiviour (see \cite{KM} for details).
The exact answer for the partition function $\log{\cal T} = \log{\cal T}_0
+ \log{\cal T}_{\theta} \equiv {\cal F}+ \log{\cal T}_{\theta}$
should also include the deformation of the oscillating
part, corresponding to the {\it massive} excitations.

An advangate of the language of the integrable systems is that it
allows one at least in principle to use a strict (in many interesting
cases explicit) formulation of the hypothetical properties of quantum field
and string theories where the basic one is given by the already mentioned
duality.

In all interesting integrable models the Liouville torus
of an integrable system is a real section of a {\it
complex} torus being for KP/Toda-theories  a Jacobian of a
(spectral) Riemann curve $\Sigma $. It is clear that there exists several
possibilities to choose a real section for the same Jacobian -- these different
choices correspond to {\it a priori} different integrable systems which
however may be related in a simple way having the same or related spectral
curves
\footnote{For example the Jacobi map might not be able to distinguish the
curves $\Sigma $ and $\tilde\Sigma $ if $\Sigma $ is a cover of
$\tilde\Sigma $.}.
Such sort of duality imposed by the ''exchange" of different cycles on the
same Jacobian (or the spectral curve itself) is defined globally
for the integrable system. Exchange between different sections of Jacobian
(different trajectories) corresponds to the exchange between the particles
and "collective excitations" (like monopoles) in field theory (see
\cite{Vafa} for the stringy explanation of this effect).

Another sort of duality works only
{\it locally} in moduli space for a completely integrable system defined by
a set of Hamiltonians $\{ h_k \}$ commuting with respect the Poisson bracket
determined by (\ref{wsform}) $\{ h_k, h_l \} = 0$. In many physical cases the
phase space has a structure of a cotangent bundle to a configuration space,
then in addition to hamiltonians one can find another distinguished set
of Poisson-commuting variables \cite{Fock}, for example, the co-ordinates on
configuration space $\{ q_k\}$:  $\{ q_k, q_l \} = 0$.
Again, the dual transformation preserves the symplectic form (\ref{wsform}).
However, now the "integrals of motion" are no longer constants on
trajectories and one can study the variation of the partition function with
respect to moduli ($h_k$ or $a_k$ variables). This gives rise to the Whitham
deformations of the finite-gap solutions, producing the exact answer for the
{\it whole} theory.

\bigskip\bigskip
I am grateful to V.Fock, A.Gorsky, I.Krichever, A.Mironov, A.Morozov and
N.Nekrasov
for the illuminating discussions and to the organizers of the Conference
for nice hospitality in Alushta. The work was partially supported by the
RFFI grant 96-02-19085 and INTAS grant 93-2058.

\end{document}